%% file: entropies.tex
\documentclass{llncs}
\usepackage{amsmath,amsfonts,pifont,amssymb,latexsym}
\usepackage{algorithm,algorithmic}
\usepackage{comment}
\usepackage{bbm}
\usepackage{stmaryrd}
\newcommand{\M}{{\mathbbm Z^d}}
\newcommand{\Z}{\mathbbm Z}
\newcommand{\N}{\mathbbm N}
\newcommand{\Ns}{{\mathbbm N_1}}
\newcommand{\R}{\mathbbm R}
\newcommand{\am}{A^{\Z^d}}

\newcommand{\deux}{\mathbbm2}

\newcommand{\sett}[2]{\left\{\left.#1\vphantom{#2}\right|#2\right\}}
\newcommand{\set}[3]{\sett{#1\in #2}{#3}}
\newcommand{\co}[2]{\left\llbracket #1,#2\right\llbracket}
\newcommand{\cc}[2]{\left\llbracket #1,#2\right\rrbracket}

\newcommand{\dfn}[1]{\textbf{#1}}
\newcommand{\ipart}[1]{\left\lfloor #1\right\rfloor}
\newcommand{\spart}[1]{\left\lceil #1\right\rceil}
\newcommand{\card}[1]{\left|#1\right|}
\newcommand{\abs}[1]{\left|#1\right|}
\newcommand{\occ}[2]{\left|#1\right|_{#2}}

\newcommand{\restr}[1]{_{\left|#1\right.}}

\newcommand{\dentr}[2]{\mathcal H_{\vec{#2}}\left(#1\right)}
\newcommand{\entr}[1]{\mathcal H(#1)}

\newcommand{\ball}[2][]{\mathcal B_{#1}(#2)}
\newcommand{\pb}[2]{\begin{quote}{\normalfont\textbf{Input:}} #1.\\{\normalfont\textbf{Output:}} #2?\end{quote}}
\newcommand{\TODO}[2][]{BLABLA #1 $\Big\langle$#2$\Big\rangle$}

\newcommand{\resp}[1]{\ (resp. #1)}
\newcommand{\ie}{, \textit{i.e.}, }
\newcommand{\etc}{, \textit{etc.}}
\newcommand{\start}[2][]{\begin{#2}[#1]\hfill\begin{itemize}}
\newcommand{\finish}[1]{\popQED\popQED\end{itemize}\end{#1}}

\title{Densities and entropies in cellular automata\thanks{
This project was supported by the Academy of Finland Grant 131558. The second author was supported by the Finnish Academy of Science and Letters and the Turku Center for Computer Science. The first author was supported by the ANR Projet Blanc ''EMC".
}}
\author{Pierre Guillon\inst2\fnmsep\inst1
\and Charalampos Zinoviadis\inst1}
\authorrunning{P. Guillon \and Ch. Zinoviadis}
\institute{Department of Mathematics, University of Turku, FI-20014 Turku, Finland,
\email{chzino@utu.fi}
\and CNRS \& Institut de Math\'ematiques de Luminy, Campus de Luminy, Case 907, 13288 Marseille cedex 9, France,
\email{pguillon@math.cnrs.fr}}

\begin{document}
\maketitle
\begin{abstract}\noindent
Following work by Hochman and Meyerovitch on multidimensional SFT, we give computability-theoretic characterizations of the real numbers that can appear as the topological entropies of one-dimensional and two-dimensional cellular automata.
\end{abstract}

\section*{Introduction}

Cellular automata are a widely-used model for complex systems or computation, consisting in a network of cells each of whose is in one among a finite number of states, that is updated synchronously in parallel as a function of the sates of its neighbors. Their entropy is a measure of how complex or random the local long-term behavior can look like.
The entropy of cellular automata has been proven uncomputable in \cite{entrind} (see also \cite{entrss} for subshifts), but the question remained whether the entropy of a single given cellular automaton could be an uncomputable number.
Recently, M.~Hochman and T.~Meyerovitch have characterized the entropies of 2-dimensional SFT \cite{entrsft} and 3-dimensional CA \cite{projsft} as, respectively, the right-computable numbers and the limits of computable increasing sequences of such numbers.
We prove here that these two classes still characterize the possible entropies of, respectively, 1-dimensional and 2-dimensional CA.
To do so, we adapt their homogeneous encoding \cite{entrsft}, J.~Kari's determinization signals \cite{nilpind} and P.~G\'acs's self-similar construction \cite{gacs}.
The result brings new equivalences between classes that are equally natural in computability theory and dynamical systems; we also believe that the construction in itself is promissing, and could help understand the real computational power of these natural models.

In Section \ref{s:prelim} we introduce the notions and a brief state of the art. In Section \ref{s:result} we state our main results, and the following sections are devoted to sketching their proofs. The algorithmic part, as well as entropy proofs and a sketch of the main construction can be found in \cite{appendix}.

\section{Preliminaries}\label{s:prelim}
\subsection{Configurations}

$\N$ will denote the set of natural numbers, $\Ns$ the set $\N\setminus\{0\}$ of positive natural numbers and $\cc ij$ the integer interval $\{i,\ldots j\}$, for $0\le i\le j$. $\R_+$ is the set of nonnegative real numbers.

\newcommand{\dinf}[1]{\vphantom{#1}^\infty{#1}^\infty}
\newcommand{\uinf}[1]{{#1}^\infty}
Let $A$ be a finite set called the \dfn{alphabet} and $d\in\Ns$ the \dfn{dimension}. Any element $x$ of $A^{\Z^d}$ is called a \dfn{configuration}, and $x_i$ is called the \dfn{state} of \dfn{cell} $i$.
The set of configurations forms a compact topological space when endowed with the product of the discrete topology.

For any $q\in A$, $\dinf q$ denotes the \dfn{$q$-uniform} configuration of $A^{\Z^d}$, all of whose cells are in state $q$. 
If $U\subset\Z^d$, $x\restr U$ is the \dfn{pattern} representing the restriction of $x$ to $U$.
For instance, we can define the central pattern $x\restr{\ball r}$ of width $r$, where $\ball r=\cc{-r}r^d$. 

\newcommand\length[1]{\left|#1\right|}

\subsection{Symbolic dynamics}
\newcommand{\finie}{\mathop\subset\limits_{\text{\tiny finite}}}

$\Z^d$ acts on $A^{\Z^d}$ by the \dfn{shift}: to any $k\in\Z^d$ we associate the homeomorphism $\sigma^k:A^{\Z^d}\to A^{\Z^d}$ defined by
$
\forall x \in \am, \forall i \in \Z^d,\sigma^k(x)_i=x_{i+k}
$. 
A ($d$-dimensional, or $d$D) \dfn{subshift} is the set $X=\set x{A^{\Z^d}}{\forall U\finie\Z^d,k\in\Z^d,\sigma^k(x)\restr U\notin\mathcal F}$
of configurations that avoid some particular set $\mathcal F$ of finite patterns. 
Equivalently, a subshift is a subset which is invariant by $\sigma^k$ for any $k\in\M$ and topologically closed.
It is of \dfn{finite type} (SFT) if $\mathcal F$ can be chosen finite.

\newcommand{\lang}{\mathcal L}
\renewcommand{\k}{\mathcal K}
Let $X\subset A^{\Z^d}$ be a subshift. 
The \dfn{language} of support $U\subset\Z^d$ of $X$ is $\lang_U(X)=\sett{x\restr U}{x\in X}$. Its \dfn{complexity} of support $U$ is $\k_U(X)=\card{\lang_U(X)}$.
The (topological) \dfn{entropy} of $X$ is $\entr X=\lim_{r\to\infty}\frac{\log{\k_{\ball r}(X)}}{\card{\ball r}}$. This is always a limit, but may be infinite.
Note that if $Y\subset B^{\Z^d}$ is another subshift, then $X\times Y$ can be essentially seen as a subshift of $(A\times B)^{\Z^d}$, and its entropy is the sum of those of $X$ and of $Y$.

A subshift $Y\subset B^\M$ is a \dfn{letter factor} of $X\subset\am$ if there exists some \dfn{letter projection} $\pi:A\to B$ such that the corresponding global map $\Pi:X\to Y$,
defined by the parallel application of $\pi$,
is onto (we say that $X$ \dfn{letter-factors} onto $Y$).
A subshift is called \dfn{sofic} if it is a letter factor of some SFT.

The same definitions hold for (one-sided) subshifts over $A^\Ns$. 

The \dfn{trace} of $X$ according to vector $\vec v\in\Z^d$ and
width $k$ is the $(d-1)$D subshift $\tau^k_{\vec v}(X)=\sett{(x\restr{\co0k\times\{0\}+n\vec v})_{n\in\Z}}{x\in X}$ over alphabet $A^k$.
\newcommand{\n}{\mathcal N}
The \dfn{directional entropy} according to vector $\vec v$ is the limit $\dentr Xv$ of the entropies of $\tau^k_{\vec v}(X)$, when $k$ goes to infinity (see \cite{milnor}). 
One can see that $\dentr X{e_2}=\lim_{k\to\infty}\lim_{r\to\infty}\frac{\log{\n_{k,r}(X)}}{r}$, where $\n_{k,r}(X)=\k_{\co0k\times\co0r}(X)$.

Let $X$ and $Y$ be 2D subshifts.
We say that $X$ \dfn{simulates} $Y$ with parameters $B,T$ if there exists $Z\subset X$ such that $X=\bigcup_{0\le i<B,0\le j<T}\sigma^{(i,j)}(Z)$ and that $Z^{<B\times T>}=\sett{(x\restr{\co{kB}{(k+1)B}\times\co{lT}{(l+1)T}})_{k,l\in\Z}}{(x_{i,j})_{i,j\in\Z}\in Z}$ is a subshift that letter-factors onto $Y$\ie any configuration of $X$ can be divided into $B\times T$ rectangles that project onto letters of $Y$.
A simulation is an \dfn{$r$-simulation} if the letters onto which an array of $(2r+1)$ horizontally consecutive rectangles of size $B\times T$ project uniquely determine the central rectangle.

The following lemma will be useful in the sequel. $(\vec{e_1},\vec{e_2})$ denotes the canonical base for $\Z^2$.
\begin{lemma}[\cite{appendix}]\label{l:simulation}
Let $X$ and $Y$ be 2D subshifts such that $X$ $l$-simulates $Y$ with parameters $B,T$.
Then, $\dentr{X}{e_1}\le \dentr{Y}{e_1}/B$ and $\dentr{X}{e_2}\le \dentr{Y}{e_2}/T$.
\end{lemma}

\subsection{Cellular automata and determinism}
A \dfn{cellular automaton} (CA) is a system $F:A^{\Z^d}\to A^{\Z^d}$ such that $F\sigma^k=\sigma^kF$; equivalently there is a \dfn{radius} $r\in\Ns$ and a \dfn{local rule} $f:A^{\ball r}\to A$ such that $\forall x\in\am,\forall i\in\M,F(x)_i=f(x\restr{i+V})$.
The \dfn{entropy} $\entr{F}$ of $F$ is the limit, when $r$ goes to infinity, of the entropy of the subshift $\sett{(F^t(x)\restr{\ball r})_{t\in\N}}{x\in\am}$.

We say that an SFT $X\subset A^{\Z^2}$ is \dfn{south-deterministic} if there is a map $F:\tau^1_{\vec{e_1}}\to\tau^1_{\vec{e_1}}$ that maps any line of a valid tiling to a unique line that can appear above\ie $\forall x\in X,j\in\Z,F(x_{\Z\times\{j\}})=x_{\Z\times\{j+1\}}$.
It is known that $F$ can actually be taken to be the restriction of a CA over alphabet $A\sqcup\{\bot\}$, where $\bot$ must be understood as ``extension not defined"; and the entropy of $F$ is equal to $\dentr X{e_1}$ (intuitively, this comes from the fact that state $\bot$ will remain forever and not contribute to the entropy).
$X$ is \dfn{south-west-deterministic} if there is the same kind of CA on the diagonal\ie $\forall x\in X,j\in\Z,F((x_{i,j})_{i=-j})=(x_{i,j})_{i=1-j}$.



Let us say that a 2D subshift 
is \dfn{S0-sofic} if it is a letter-factor of some south-deterministic SFT with null entropies\ie directional entropy $0$ according to any vector.






\subsection{Effectiveness}
\newcommand\m{\mathcal M}
In $A^{\Z^d}$, it is easy to enumerate computationally a base of open sets (consider the sets of configurations sharing a given pattern as a central pattern). That way, we can define an \dfn{effectively closed} subset $S\subset A^{\Z^d}$ as the complement of the union of a computable sequence of open sets. It is an \dfn{effective subshift} if, besides, it is a subshift.
For instance, trace of SFT are effective subshifts.
Effectively closed sets can also be defined in other Cantor sets; in $A^{\Ns}$ they correspond to sets of configurations that are not ultimately rejected when scanned by some given TM.
An \dfn{effective system} is an effectively closed subset $S\subset (A^{\Ns})^{\Z^d}$ which is invariant by the $\Z^d$-shift. Intuitively, it is a dynamical system where the preimages of open sets can be computed.

A \dfn{$\Pi_1$} (or right-computable) number is the limit of a decreasing computable sequence of rational numbers.
A \dfn{$\Sigma_2$} number is the limit of an increasing computable sequence of $\Pi_1$ numbers.
Equivalently, there exists an algorithm that on input $k$ outputs the code of another algorithm $\m_k$ such that $\m_k$ enumerates the approximations of a $\Pi_1$ number $h_k$, the sequence $h_k$ is increasing and converges to $h$.
The set of $\Sigma_2$ is strictly larger than the set of $\Pi_1$ numbers, which, in turn, is strictly larger than the set of computable ($\Delta_1$) numbers.
We refer to \cite{rhierarchy} for more on these classes of numbers (and many more).
\begin{remark}
The binary representations of real numbers from an interval $[0,\alpha]$ form an effectively closed subset of $\deux^\N$ if and only if $\alpha$ is $\Pi_1$.
\end{remark}

\section{Results}\label{s:result}

Some evidence of the \emph{computing power} of a given model can be given by studying the class of numbers that can be realized as entropy. Elegant characterizations have recently been achieved for multidimensional SFT.
\begin{theorem}[\cite{entrsft,projsft}]
For $d\ge2$, the class of entropies of $d$-dimensional SFT\resp{$d$-dimensional sofic subshifts, effective subshifts} is $\R_+\cap\Pi_1$.
\end{theorem}
In the broader case of effective systems (and as a consequence for high-dimension CA), the class of entropies that can be realized is larger.
\begin{theorem}[\cite{projsft}]\label{t:entrca}
For $d\ge3$, the class of entropies of $d$-dimensional CA\resp{effective systems} is $\R_+\cap\Sigma_2\cup\{\infty\}$.
\end{theorem}

The last two theorems have left open the case of entropies realized by 1D and 2D CA, that are both included in $\Sigma_2$.
The main purpose of the present article is to solve these two remaining cases.
The first step of the answer is given by the following result:
\begin{theorem}[\cite{coneentr}]\label{t:entrcone}
The entropy of a 1D CA 
is equal 
to the entropy of some trace of the corresponding 2D SFT. 
\end{theorem}
From the theorem above, 
the entropy of a 1D CA is thus $\Pi_1$.
We will actually prove that the converse is also true.
\begin{theorem}\label{t:entr1d}
The class of entropies of 
1D CA is $\R_+\cap\Pi_1$.
\end{theorem}
This class of numbers is thus strictly weaker than the possible entropies of 3D CA, characterized in \cite{projsft}. However, this is not true for the 2D case.
\begin{theorem}\label{t:entr2d}
The class of entropies of 2D CA is $\R_+\cap\Sigma_2\cup\{\infty\}$.
\end{theorem}

\section{Construction}\label{s:proof}
\subsection{Density encoding}\label{ss:dens}
This subsection is devoted to encoding data in the density of the configurations. 
The most relevant is actually the binary case, which follows the construction in \cite{entrsft}. 

A \dfn{$1$-net} is a family $(2^n\Z+k_n)_{n\in\Ns}$ of pairwise disjoint subsets of $\Z$ called \dfn{levels}, where $(k_n)_{n\in\Ns}\in\Z^\Ns$.
It can be seen that for any $1$-net, there is at most one cell $i\in\Z$ which does not belong to any level.

\newcommand\ans{\widetilde{A^\Ns}}
Let us denote $\occ ua$ the number of occurrences of letter $a$ in word $u$.
The \dfn{frequency} of a letter $a\in A$ in some one-dimensional configuration $x\in A^\Z$ is, if ever it exists, the limit
$\delta_a(x)=\lim_{r\to\infty}{\occ{x\restr{\ball r}}a}/{\card{\ball r}}$.

If $\alpha,\beta\in A^\Ns$, we note $\alpha\sim\beta$ if $\alpha=\beta$ or there exists $i\in\Ns$ such that $\forall j<i,\alpha_j=\beta_j$, and $\forall j>i,\alpha_i=\beta_j$ and $\alpha_j=\beta_i$. This is an equivalence relation, for which all the classes have cardinal one or two. As an example, two binary sequences are equivalent for $\sim$ if and only if they represent binary expansions of the same real number in $[0,1[$. Let $\ans$ be the quotient of $A^\Ns$ by this equivalence relation. It can be endowed with the induced topology from the product topology. We will often confuse a sequence $x$ and its equivalence class.

\newcommand\D{\mathcal D}
\newcommand\decod{\m_\triangleright}
If $\alpha\in A^\Ns$, we note $\D_\alpha\subset A^\Z$ the set of \emph{T\oe plitz} configurations which are constantly equal to $\alpha_n$ on level $2^n\Z+k_n$ for some $1$-net $(2^n\Z+k_n)_{n\in\N}$.
If $S\subset A^\Ns$, we note $\D_S=\bigcup_{\alpha\in S}\D_\alpha$.
These sets have interesting properties.
\begin{remark}\label{r:dens}~\begin{enumerate}
\item For any nonempty closed set $S\subset A^\Ns$, $\D_S$ is a nonempty subshift.
\item The frequency of any letter $a\in A$ in any configuration $x\in\D_\alpha$ is $\sum_{\alpha_i=a}2^{-i}$.\\
In particular if $\alpha$ is binary, then it is a binary expansion of $\delta_1(\alpha)$.
\item\label{i:sim} If $\alpha\sim\beta$, then $\D_\alpha=\D_\beta$;
otherwise, $\D_\alpha\cap\D_\beta=\emptyset$.
\item\label{i:odd} Let $x\in\D_\alpha$, $j\in\Z$, and $i$ be an odd number.
Then $x\restr{i\Z+j}$ is still in $\D_\alpha$.
\end{enumerate}\end{remark}
Point \ref{i:sim} of the previous remark suggests that it is relevant to talk about $\D_\alpha$\resp{$\D_S$} for an equivalence class $\alpha\in\ans$, or for a real number $\alpha\in[0,1]$\resp{a set $S\subset\ans$ of classes}.

Moreover, the sequence $\alpha$ encoded in the densities of the subshift can actually (up to equivalence) be effectively approximated by reading finite patterns.
\begin{lemma}[\cite{appendix}]\label{l:dens}
There exists a TM $\decod$ which, given a word $u$ over alphabet $A$, outputs a word $v$ such that, if $u=x\restr{\co0{2^n}}$ for some $x\in\D_\alpha$ and some $n\in\N$
, then $v=\beta_{\cc1n}$ for some $\beta\sim\alpha'$ and $\alpha'_{\cc1n}=\alpha_{\cc1n}$.
\end{lemma}

We say that a TM has input in $\ans$ if it reads sequences of $A^\Ns$ as input, and gives the same result for sequences in the same equivalence class. We can also assume that, if $\alpha\sim\alpha'$, then this TM stops after the same number of steps for $\alpha$ and $\alpha'$. 
\begin{lemma}[\cite{appendix}]\label{l:denscode}
For any TM $\tilde\m$ with input in $\ans$, there exists a TM $\m$ with input in $A^\Ns$ such that:
\begin{itemize}
\item If $\tilde\m$ halts over input $\alpha\in A^\Ns$, then there exists $k\in\N$ such that for 
any configuration $x\in\D_\alpha$, $\m$ halts over input $x\restr{\co0k}$ before time $k$;
\item otherwise, 
$\m$ does not halt over any input $x\in\D_\alpha$. 
\end{itemize}
\end{lemma}

The following corollary is a direct application of Lemma \ref{l:denscode} with a machine rejecting configurations outside some effectively closed set.
\begin{corollary}\label{c:denscode}
If $S\subset\ans$ is effectively closed, then $\D_S$ is an effective subshift.
\end{corollary}

\subsection{Checking homogeneity}
Our proof involves a deterministic SFT
that is built layer by layer: the state of each cell is in a product of alphabets that we define one after the other, each layer having to respect some local constraints in how it can be superimposed with the previous ones. 
For $\alpha\in A^\Ns$\resp{$S\subset A^\Ns$}, let us note $\D_\alpha^*$\resp{$\D_S^*$} the set of configurations $x\subset A^{\Z^2}$ which are constant vertically, and where each row $(x_{i,k})_{i\in\Z}$ is in $\D_S$, for $k\in\Z$.

The purpose of this subsection is to build an SFT which checks that some layer is well homogeneous, in the sense of the following lemma
; this follows \cite[Section 6]{entrsft}, but contrary to this, keeping determinism and null entropies forces us to go back to the actual SFT construction rather than directly invoke Mozes's theorem for $2\times2$-substitutions. 
\begin{lemma}\label{l:homog}
$\D_{A^\Ns}^*$ is S0-sofic.
\end{lemma}
We will only give a sketch of the proof. 
A \dfn{$2$-net} is a family $(I_n\times J_n)_{n\in\Ns}$ of products of levels of two $1$-nets $(I_n)_{n\in\Ns}$ and $(J_n)_{n\in\Ns}$. Each $I_n \times J_n$ itself is called the \emph{level $n$} of the net.
The $I_n$\resp{$J_n$} being pairwise disjoint, it follows that a horizontal\resp{vertical} line can intersect at most one level of the $2$-net.
If $i \in I_n$, then $\{i\}\times \Z$ is called a \emph{column of level $n$}. By definition, columns of level $n$ appear with horizontal period $2^n$.

In \cite{robinson}, Robinson constructed an SFT $R$ in which every configuration is divided regularly into squares of size $2^n$ for every $n$. 
In particular, he mentions, in other terms, the following property about the good repartition of a particular state called a cross.
\begin{lemma}[\cite{robinson}]\label{l:net}
For every $x\in R$, the set $\set{i}{\Z^2}{x_{i}\text{ is a cross}}$ is a $2$-net.
\end{lemma}

Now, this SFT has been made deterministic in \cite{nilpind}, by adding to it a layer with signals that forbid some configurations that would share the same bottom-left half as another one. The result can be restated as follows.
\newcommand\rdet{\overrightarrow R}
\begin{lemma}[\cite{nilpind}]\label{l:detrob}
There exists a south-west-deterministic SFT $\rdet$ that letter-factors onto some nonempty subsystem of $R$.
\end{lemma}

\newcommand\rh{\tilde R'}
\newcommand\rhs{\tilde R}
\begin{proof}[of Lemma \ref{l:homog}]
Let us first define a south-west-deterministic SFT $\rh$, in which configurations are vertically constant and correspond horizontally to $\D_{A^\Ns}^*$.
$\rh\subset R\times A^{\Z^2}\times A^{\Z^2}$ is defined with three layers:
the first one contains the deterministic Robinson SFT $\rdet$; the second one is constant horizontally; the third one is constant vertically. We additionally require that if the first layer is a cross, then the other two must coincide. $\rh$ is south-west-deterministic, since all three of its layers are.
Now it is not difficult to turn this SFT into a south-deterministic one, by simply considering $\rhs=\sett{(x_{(i,j-i)})_{(i,j)\in\Z^2}}{(x_{i,j})_{(i,j)\in\Z^2}\in\rh}$, whose columns correspond to columns of $\rh$, but lines correspond to north-west-to-south-east diagonals of $\rh$.

Null entropies come from the substitutive nature of $R$, which is transmitted to $\rhs$.
More details about this can be found in \cite{appendix}.
\qed\end{proof}

\subsection{Checking the density}

\input{self-similar}

\section{The second dimension}\label{s:2d}
Let us now prove Theorem \ref{t:entr2d}, dealing with 2D CA. The first inclusion is direct from Theorem \ref{t:entrca}.
The idea here will be to realize, in each horizontal slice, some right-computable number, as in the previous section. These slices will actually be parameterized by some index encoded in its density, that is increased by one between consecutive slices, and that will give a sequence approximating the wanted $\Sigma_2$. The trick is that the encoding has to be spare in order to prevent limit configurations to achieve too much entropy; this has to be compensated by having actual groups of consecutive slices hold the same parameter.

Let us denote by $(w)_4$ the 4-ary representation of a natural number $w\in\N$ over $\{0^{\prime},1^{\prime},2^{\prime},3^{\prime}\}$.
Let $(S_k)_{k\in\Ns}$ be a computable sequence of effectively closed subsets of $\ans$,
and $S=\sett{*^k(w)_4y}{k \in \N, 0 \le w \le 4^k-1, y\in \D_{S_k}} \cup \{\uinf*,\uinf\sharp\}$ a set of sequences over alphabet $A=\{*,0^\prime,1^\prime,2^{\prime},3^{\prime},0,1,\sharp\}$.
Consider $S^{\prime}=S_1\cup S_2\cup S_3$, where:
\begin{eqnarray*}
S_1&=&\set{(z,z')}{S^2}{\exists k,y,w\in\co0{4^k-1},z=*^k(w)_4y\text{ and }z'=*^k(w+1)_4y};\\
S_2&=&\set{(z,z')}{S^2}{\exists k,y,z=*^k(4^k-1)_4y\text{ and }z'=\uinf\sharp};\\
S_3&=&\set{(z,z')}{S^2}{z=\uinf*\text{ and }z'=\uinf*\text{ or }\exists k,y,z'=*^k(0)_4y}.
\end{eqnarray*}

\begin{lemma}[\cite{appendix}]\label{l:2dsecond}
$S^{\prime}$ is an effectively closed subset of $\widetilde{(A\times A)^{\Ns}}$.
\end{lemma}

We are now ready to characterize the entropies of 2D CA.
Similarly to the one-dimensional case, a 2D CA corresponds to a south-deterministic 3D SFT, up to adding a spreading state, and its entropy can be seen as the directional one for the south-to-north unitary vector.
\begin{proof}[of Theorem \ref{t:entr2d}]
Let $\alpha_k$ be a computable sequence of $\Pi_1$ numbers, $S_k=[0,\alpha_k]$, $\m$ the TM given by Lemma~\ref{l:2dsecond}, $Y$ the 2D SFT given by Lemma \ref{l:effs0}.

Consider now the following 3D SFT $Y'$: each horizontal slice must satisfy the conditions of $Y$.
The only vertical local constraint we add is the following:
the second letter (in $A$) of the pair held by a tile is equal to the first letter of the tile on top of it.
Intuitively, the way to think about this is that when a horizontal slice is considering whether it should accept or reject its input (the first sequence it holds), it can also read as input the sequence of the slice above it (the second sequence).

$Y'$ is south-deterministic. Indeed, every horizontal slice is an element of $Y$, which is a 2D south-deterministic SFT. Hence, if we know a  slice $x\restr{\Z \times\{n\} \times \Z}$, we can uniquely determine $x\restr{\Z \times\{n+1\}\times\Z}$.
Moreover, $Y'$ has null entropies, as a subshift of an infinite product of 2D SFT with null entropies.

Let us now modify the SFT in order to get the wanted entropy. We need to understand the structure of the configurations.
From now on, we forget the second sequence encoded in every horizontal slice and we work only with the first one. If $z_k$ is the sequence encoded in the $k$th horizontal slice, then the sequence $(z_k)_{k \in \Z}$ can only have one of the following forms:
\begin{itemize}
\item $z_k = \uinf*$, for all $k \in \Z$;
\item there exist $m \in\Z, k \in \Ns$ and $y \in \D_{[0,\alpha_k]}$ such that $z_i = \uinf*$ for $i<m$, $z_i = *^k(i-m)_4y$ for $m\le i<m+4^k$, and $z_{i} = \uinf\sharp$ for $i \geq m+4^k$.
\item $z_k = \uinf\sharp$, for all $k \in \Z$;
\end{itemize}
This follows directly from the definition of $S^{\prime}$. For $k \in \Ns$, let $Y^{\prime}(k) \subseteq Y^{\prime}$ consist of those configurations whose horizontal slices are either $\uinf*$, $\uinf\sharp$, or contain $*^k(0)_4y$ for some $y\in\D_{[0,\alpha_k]}$. It is a subshift. 

Let us allow splitting of the letter $1$ into two (by adding a second, binary, layer, as in the proof of Theorem \ref{t:entr1d}), independently in every horizontal slice. Then, in configurations of the subsystem $Y^{\prime}(k)$ there are $4^k$ slices where splitting is done and each one contributes up to $4^{-k}\beta$ to the entropy, where $\beta\in[0,\alpha_k]$ is such that $y\in\beta$. This happens because in every slice, $y$ is encoded in $2$-net starting from level $2k$. Since splitting is done independently in $4^k$ slices, the entropy of the subsystem $Y^{\prime}(k)$ is $\beta$. By the variational principle, and since the nonwandering system of the CA is included in the disjoint union of the $Y^{\prime}(k)$ and the trivial subsystems, we have that the entropy of $F$ in the vertical direction is:
\[\entr{F} = \sup_{k\in\N}\sup_{0\le\beta\le\alpha_k}\beta=\sup_{k\in\N}\alpha_k,\]
which is the wanted $\Sigma_2$ number.
\qed\end{proof}

\section*{Conclusion}
We have reached a characterization of the entropies of CA in terms of computability classes. This is inspired by what had been done over multidimensional SFT, but the construction presents some intrinsically interesting points, such as determinization widgets, self-similar construction, or a generalized encoding of configurations into densities.

This problem helps us understand what kind of results on tilings could be adapted to CA, that is when one of the dimensions of the system actually represents a deterministic temporal evolution.
It could be interesting to try to adapt some more results from multidimensional symbolic dynamics, such as the substitutions of \cite{mozes}, or the characterization of subactions in \cite{projsft,fixpttile}. 
 Nevertheless, when translating into cellular automata, we will in general have to deal with wandering points, which could be omitted here in the study of entropy but may sometimes alter significantly the results.

Among open problems, we could try to characterize the entropies of restricted classes of CA: requiring transitivity constraints, or reversibility.
The latter case might be achieved by adapting our proof while requiring two-way determinism in the underlying tilings (but again extending it to a full set of configurations may be difficult).
We could also study the entropies of other computationally-inspired dynamical systems, such as Turing machines with moving tapes. 

\bibliographystyle{splncs}
\bibliography{entropies}

\newpage\appendix
\input{tilemacros}

\end{document}

%% file: self-similar.tex
%
%
\newcommand{\addr}{\texttt{Addr}}
\newcommand{\age}{\texttt{Age}}
\newcommand{\info}{\texttt{Info}}
\newcommand{\rmail}{\texttt{Rmail}}
\newcommand{\prog}{\texttt{Prog}}
\newcommand{\work}{\texttt{Work}}
\newcommand{\lev}{\texttt{Level}}
\newcommand{\lmail}{\texttt{Lmail}}
\newcommand{\newinfo}{\texttt{NewInfo}}
\newcommand{\checkk}{\texttt{Check}}
\newcommand{\checkkk}{\texttt{Check$_2$}}
\newcommand\yn{Y_n^{\prime}}

In this section, we 
construct a south-deterministic SFT with null entropies which letter-factors onto $\D^*_S$.
In the SFT, there is a special layer which consists exactly in $\D_S^*$: from Lemma \ref{l:homog}, we can a priori assume that all configurations of this layers are in $\D_{A^\Ns}^*$, by implicitly having a layer in $\rhs$.
We will now add a layer whose purpose is to check that if $x\in\D_\alpha^*$ is read from this layer, with $\alpha\in\ans$, then $\alpha$ is really in the wanted set $S$, by simulating the application of a machine $\m$ corresponding to the machine $\tilde\m$ that rejects any configuration that is not in $S$ (see Lemma \ref{l:denscode}).

A naive simulation of the machine for an infinite time would create invalid limit configurations.
A solution to this problem is to build the additional layer in a self-similar way, in the fashion of \cite{gacs,fixpttile,nexpdir}: we build a family of south-deterministic SFT $(Y_n)$ such that $Y_n$ simulates the TM for $n$ steps, and also simulates $Y_{n+1}$ with some parameters $B_n,T_n$. That way, if $n$ was not enough to figure out that the input had to be rejected, then a higher level will notice it. More precisely, $Y_n$ will be able to apply the TM over the input $x\restr{B_{n+1}\co0{B_n}+j}$ for some $j\in\co0{B_{n+1}}$.
The simulation of $Y_{n+1}$, as defined previously, consists in dividing naturally 
every valid configuration of $Y_n$ into rectangles of size $B_n \times T_n$ 
called the \dfn{$Y_n$-macrotiles}.
An important feature is that this family admits a uniform description: one single SFT is actually described. Each configuration is conscious of the level $Y_n$ it belongs to, and will check that it simulates a configuration of the next one.
The details of the construction ensuring these conditions can be found in \cite{appendix}.

The following lemma 
applies machine $\m$ from Lemma \ref{l:denscode} to finite configurations composed of some arithmetic progressions in lines of the SFT, that are still in $\D_\alpha$. Null entropies come from the self-simulation.
\begin{lemma}\label{l:effs0}
If $S\subset\ans$ is an effectively closed set, then $\D_S^*$ is S0-sofic.
\end{lemma}

\subsection{From density to entropy}
Finally, let us see how Lemma \ref{l:effs0} can be used to prove Theorem \ref{t:entr1d}: it simply independently splits each letter $1$ into two letters, so that its density is transformed into entropy.
\begin{proof}[of Theorem \ref{t:entr1d}]
One direction corresponds to Theorem \ref{t:entrcone}.
Let us prove the converse.
	Should we make the product with the shift over $2^{\ipart\alpha}$ symbols, whose entropy is $\ipart\alpha$, we can assume that $\alpha\in[0,1[$.
	
\newcommand{\dsplit}{\D_S^\nabla}
Let $F$ be the shift composed with the CA corresponding to the deterministic SFT given by Lemma \ref{l:effs0} 
for the effectively closed set $S$ consisting of binary representations of real numbers from the interval $[0,\alpha]$, $A$ its alphabet, and $\pi:A\to\deux$ be the corresponding letter projection. 
Let $\tilde F$ be the CA over alphabet $(A \times\{0\})\sqcup(\pi^{-1}(1)\times\{1\})$ 
 such that the first component performs $F$ and the second one performs the shift.
\ie
in the first component we can see the $0$-entropy $F$ and, in the second one the one-dimensional subshift:
\[
\dsplit=\set{(y_i)_{i\in\Z}}{\deux^\Z}{\exists(x_i)_{i\in\Z}\in\D_S,\forall i\in\Z, \text{ if } x_i=0, \text{ then } y_i=0}.
\]
It is known that the entropy of a product is the sum of the entropies, hence the entropy of $\tilde F$ is that of $\dsplit$.

$\k_U(\dsplit)=\sum_{u\in\lang_U(\D_S)}2^{\occ u1}$ can be bounded by $\k_U(\D_S)2^{\sup_{u\in\lang_U(\D_S)}\occ u1}$. Hence, the entropy $\entr\dsplit$ is:
\[\entr\dsplit=\lim_{r\to\infty}\frac{\log\k_{\ball r}(\dsplit)}{\card{\ball r}}\le\entr{\D_S}+\lim_{r\to\infty}\sup_{u\in\lang_{\ball r}(\D_S)}\frac{\occ u1}{\card{\ball r}}.\]
However, since $\entr{\D_S} = 0$, $\entr\dsplit$ is not more than the maximal density $\alpha$ of configurations of $\D_S$.
Conversely, if $x\in\D_\alpha\subset\D_S$, then $\lang_{\ball r}(\dsplit)\supset\sett{(x_i,y_i)_{\abs i<r}}{\forall i\in\Z,y_i\in\{x_i,2x_i\}}$; hence $\k_{\ball r}(\D_S)\ge2^{\occ{x_{\ball r}}1}$ and $\entr\dsplit\ge\limsup_{r\to\infty}\occ{x_{\ball r}}1=\alpha$.
Therefore, $\entr {\tilde F}=\entr\dsplit=\alpha$.
\qed\end{proof}

%% file: tilemacros.tex
\section{Algorithms}
\begin{proof}[of Lemma \ref{l:dens}]
Consider 
the following algorithm: 
\begin{algorithmic}
\REQUIRE A word $u=u_1\ldots u_{\length u}$ over alphabet $A$.
\IF{$\length u=1$}
\RETURN the empty word.
\ENDIF
\IF{$\exists a\in A,I\subset\cc1{\length u},\card I=\length u/2,\forall i\in I,u_i=a$}
\RETURN $a$ concatenated to the word returned by this same algorithm applied to $(u_i)_{i\notin I}$.
\ENDIF
\RETURN Error.
\end{algorithmic}
Let $(k_j)_{j\in\Ns}$ and $\alpha\in A^{\Ns}$ be such that $x_i=\alpha_j$ for any $i\in2^j\Z+k_j$ and any $j\in\Ns$.
The definition of 
$1$-net gives, for any $j\le n$, $\card{\co0{2^n}\cap(2^j\Z+k_j)}=2^{n-j}$. The levels being disjoint, we get $\card{\co0{2^n}\setminus\bigcup_{j\le n}(2^j\Z+k_j)}=2^n-\sum_{j\le n}2^{n-j}=1$.
Besides, the number of occurrences of any letter $a\in A$ in $u=x\restr{\co0{2^n}}$ is
\[\occ ua=\sum_{j\le n,\alpha_j=a}\card{\co0{2^n}\cap(2^j\Z+k_j)}+\card{\set i{\co0{2^n}\setminus\bigcup_{j\le n}(2^j\Z+k_j)}{x_i=a}},\]
which is equal, with a difference of at most $1$, to $\sum_{j\le n,\alpha_j=a}2^{n-j}$.
In particular, $\alpha_1$ occurs at least $2^{n-1}$ times.
If the algorithm choses $\gamma(u)_1=\alpha_1$, then the statement is obtained by recurrence on the logarithm of the length of $u$.
Now if the algorithm choses $\gamma(u)_1\ne\alpha_1$, it means that these two letters were both equally present (each covering half of $\co0{2^n}$). It is not difficult to see that this is possible only if $\alpha_j=\gamma(u)_1$ for any $j\in\cc2n$, and that in this case the algorithm will output $\gamma(u)_1\alpha_1^{n-1}=\beta_{\cc1n}$ for $\beta=\gamma(u)_1\uinf{\alpha_1}\sim\alpha_1\uinf{\gamma(u)_1}$.
\qed\end{proof}

\begin{proof}[of Lemma \ref{l:denscode}]
Let $\m$ be the TM performing the following algorithm:
\begin{algorithmic}
\REQUIRE An infinite word $x\in A^\Ns$.
\FOR{$t\in\Ns$}
\STATE Apply $\decod$ (from Lemma \ref{l:dens}) to $x\restr{\co0{2^t}}$; let $v$ its output.
\STATE Perform $t$ steps of algorithm $\tilde\m$ over finite input $v$.
\ENDFOR
\end{algorithmic}
\begin{itemize}
\item Assume that $\tilde\m$ halts over input $\alpha\in A^\Ns$ after some time $t$. 
Let $n$ be the time needed to perform completely the loop $t$ of the algorithm (over some infinite input).
From the algorithms of $\m$ and $\decod$, it can be seen that this time depends only on $t$ (and $\m$, but not on the input).
Let 
$x\in\D_\alpha$ be some configuration.
At loop $t$ we have computed $v=\beta_{\cc1t}$ for some $\beta\sim\alpha'$ and $\alpha'_{\cc1t}=\alpha_{\cc1t}$, and simulated $\tilde\m$ over input $\beta$. The machine halts before $t$ steps over input $\alpha$, so it also does over input $\alpha'$, since they have the same prefix of size $t$, so it cannot make the distinction between the two at that point. From the property of the machine and the fact that $\beta\sim\alpha'$, we can conclude that it also stops in $t$ steps over input $\beta$.
\item On the contrary, assume that 
$\m$ halts over some input $x\in\D_\alpha$. This means that $\tilde\m$ halts within $t\in\Ns$ steps over some input $v$, which was computed by $\decod$, which is then equal to some $v=\beta_{\cc1t}$ for some $\beta\sim\alpha'$ and $\alpha'_{\cc1t}=\alpha_{\cc1t}$. Then it means that $\tilde\m$ halts over input $\alpha'\sim\beta$ before $t$ steps, hence over input $\alpha$.
\qed\end{itemize}
\end{proof}

\begin{lemma}\label{l:2dcode}
$S$ is an effectively closed subset of $\ans$.
\end{lemma}
\begin{proof}
We can see that $S$ respects the equivalence classes.
The following algorithm effectively rejects exactly the elements outside $S$:
\begin{algorithmic}
\REQUIRE A configuration $z=z_1z_2\ldots\in A^\Ns$.
\IF{$z_1=\sharp$}
\FOR{$k\in\Ns$}
\IF{$z_k\ne\sharp$}
\RETURN Error.
\ENDIF
\ENDFOR
\ELSE
\FOR{$k\in\Ns$}
\STATE if $z_k\ne*$, break out of the loop, and remember $k$. 
\ENDFOR
\FOR{$i\in\cc k{2k-2}$}
\IF{$z_i\notin\{0',1',2^{\prime},3^{\prime}\}$}
\RETURN Error.
\ENDIF
\ENDFOR
\STATE Apply the TM corresponding to $S_k$ to the sequence $z_{2k+1}z_{2k+2}\ldots$\\
(it should be considered as rejected if some letter $z_{2k+i}\notin\deux$).
\ENDIF
\qed\end{algorithmic}
\end{proof}

\begin{proof}[of Lemma \ref{l:2dsecond}]
It is easy to design an algorithm that rejects all the pairs of words outside of $S^{\prime}$ using the algorithm of Lemma~\ref{l:2dcode} as a subroutine.
\qed\end{proof}

\section{Null entropies constructions}
\begin{proof}[of Lemma \ref{l:simulation}]
We prove the claim only for the vertical direction, since the horizontal case is analogous.
By definition, $\dentr{X}{e_2}=\lim_{k\to\infty}\lim_{r\to\infty}{\frac{\log\n_{k,r}(X)}{r}}$. By passing to a subsequence, we can write $k=k'B$ and $r=r'T$. Then, a rectangular pattern $b$ of size $k \times r$ is contained into a pattern consisting of $(k'+1) \times (r'+1)$ rectangles of size $B \times T$, each of which corresponding to a letter of $Y$. SInce the simulation has radius $l$, if we thicken this pattern on the left and right by $l$ to obtain a $(k'+1+2l) \times (r'+1)$ pattern of Y, then the central $(k'+1) \times (r'+1)$ $B \times T$-rectangles are uniquely determined. Finally, this pattern together with the coordinates of the bottom-left corner of $b$ in the $B \times T$ rectangle that contains it uniquely determine $b$. Therefore,
\begin{eqnarray*}
\dentr{X}{e_2}&=&\lim_{k'\to \infty}\lim_{r'\to \infty}\frac{\log{\n_{k'B,r'T}(X)}}{r'T}\\
&\le&\lim_{k'\to \infty}\lim_{r'\to \infty}\frac{\log BT\n_{k'+1+l,r'+1}(Y)}{r'T}=\frac{\dentr{Y}{e_2}}T,
\end{eqnarray*}
which proves the claim.
\qed\end{proof}

\begin{proof}[of the entropy part of Lemma \ref{l:homog}]
It remains to prove that $\rh$ has null entropies. 
First of all, we prove that $\dentr{R}{e_2}=0$. This is true because there exists an SFT $N$ which letter-factors onto $R$ and $0$-simulates itself injectively with parameters $2,2$ by \cite{twobytwo}. Therefore, according to Lemma~\ref{l:simulation}, $\dentr{N}{e_2}\le\dentr{N}{e_2}/2$, which means that $\dentr{N}{e_2}=0$. Since $N$ letter-factors onto $R$, we also have that $\dentr{R}{e_2}=0$.

Adding the diagonal signals of $\rdet$ does not increase the directional entropy: it is mentioned in \cite{nilpind} that for rectangles of arbitrary size, there are only 4 cells where we have a choice for the diagonal signals. Therefore, $\n_{k,r}(\rdet)\le s^4\n_{k,r}(R)$, where $s$ is the number of diagonal signals. From the last equation, we immediately get that $\dentr{\rdet}{e_2} \le \dentr{R}{e_2}$.

Similarly, the horizontal and vertical signals of $\rh$ do not increase the directional entropy. This is true because a square of size $k \times r$ intersects at most $(\log{k}+1)+(\log{r}+1)$ different levels of the $2$-net. If we specify the letters of these levels, then all of the signals are uniquely determined. There are $2$ choices for every level, therefore $\n_{k,r}(\rh)\le 2^{\log{k}+\log{r}+2} \n_{k,r}(\rdet)\le 4kr\n_{k,r}$. Hence, $\dentr{\rh}{e_2} \le \dentr{\rdet}{e_2}$=0.

In the same way, we can prove that $\dentr{\rh}{e_1}=0$. By \cite{expsubd}, this implies that all the directional entropies of $\rh$ are equal to $0$. Since $\rhs$ is a shifted version of $\rh$, for every $\vec{v} \in \Z^2$, $\dentr{\rhs}{v} = 0$, which is what we wanted to prove.
\qed\end{proof}

\begin{lemma}\label{l:repr}
$x\in\D_\alpha^*$ is represented in $Y_1$ if and only if $\alpha\notin S$.
\end{lemma}
\begin{proof}
\begin{itemize}
\item Let $y\in Y_1$ be such that the $\D_{A^\Ns}^*$ layer of $y$ is $x\in\D_\alpha^*$. Let $\tilde x\in\D_\alpha$ denote a line of $x$ (recall that $x$ is constant vertically).
$Y_n$-macrotiles present in $y$ have applied, without halting, $n$ steps of TM $\m$ over input $\tilde x\restr{B_{n+1}\co0{B_n}+j}$ for some $j\in\co0{B_{n+1}}$. From Point \ref{i:odd} of the remark in Subsection \ref{ss:dens}, $\tilde x\restr{B_{n+1}\Z+j}\in\D_\alpha$. Thus we have, for any $n$, a configuration of $\D_\alpha$ that is not rejected by $\m$ within $n$ steps. By Lemma \ref{l:denscode}, this means that $\tilde\m$ does not halt over input $\alpha$\ie $\alpha\in S$.
\item
Assume, on the contrary, that $\tilde\m$ does not halt over $\alpha$. By Lemma \ref{l:denscode}, $\m$ never halts on any configuration $\tilde x\in\D_\alpha$ (and any $x\restr{B_{n+1}\Z+j}$); hence there exist $Y_n$ macrotiles for every $n$. Every $Y_n$ macrotile gives rise to a $B_1B_2 \cdots B_n \times T_1T_2 \cdots T_n$ rectangle validly tiled by $Y_1$. Therefore, $Y_1$ can tile arbitrarily large rectangles, which means that it can also tile the plane.
\qed\end{itemize}\end{proof}

\begin{proof}[of Lemma \ref{l:effs0}]
$Y_n$, from Lemma \ref{l:repr}, is a south-deterministic SFT which letter-factors onto $\D_S^*$.

Let us now prove that $\dentr{Y_n}{e_2}=0$ (a symmetrical argument will give that $\dentr{Y_n}{e_1}=0$, and \cite{expsubd} has us obtain all directions).
For every $n \in \Ns$, $Y_n$ $1$-simulates $Y_{n+1}$ with parameters $B_n,T_n$. On a second layer, there is a configuration of $\rhs$, which is independent from the first $Y_n$-layer.
Therefore, Lemma~\ref{l:simulation} gives that $\dentr{Y_n}{e_2} \le \frac{\dentr{Y_{n+1}}{e_2}}{T_n} + \dentr{\rhs}{e_2}= \frac{\dentr{Y_{n+1}}{e_2}}{T_n}$.
Inductively, we can show that for every $n,m \in \Ns$, $\dentr{Y_n}{e_2} \le \frac{\dentr{Y_{n+m+1}}{e_2}}{T_n \cdots T_{n+m}}$.
Also, $T_{n+m} \geq B_{n+m} \geq \log{\card{A_{n+m+1}}}$, where $A_{n+m+1}$ is the alphabet of the SFT $Y_{n+m+1}$. This implies that $\dentr{Y_{n+m+1}}{e_2}/T_{n+m} \le 1$, hence $\dentr{Y_n}{e_2} \le \frac{1}{T_n \cdots T_{n+m-1}}$, for every $m \in \Ns$. Since $T_n > 1$ for all $n$, it follows that $\dentr{Y_n}{e_2}=0$.
\qed\end{proof}

\section{Details of the macrotile construction}
\subsection{Fields of the macrotiles}

Each $Y_n$-tile contains a state divided into the following fields:
\begin{itemize}
\item[\lev:]This is equal to the unary word $1^n$.

\item[\addr, \age:]These two fields contain two integers $i\in\co0{B_n}$ and $j\in\co0{B_n}$ respectively, that correspond to the \dfn{coordinates} of the tile.
The local constraint for the coordinates are quite natural: the right neighbor of a tile with coordinates $(x,y)$ must have coordinates $(x+1 \bmod B_n,y)$ and its upper neighbor must have coordinates $(x,y+1 \bmod T_n)$.

\item[\info:]This field contains a letter from the alphabet $\Gamma=\{0,1,/,\sharp\}$.
For every $Y_n$-macrotile, the word over $\Gamma$ of length $B_n$ consisting of the \info\ fields of the tiles at positions $(i,0)_{0\le i<B_n}$, represents the complete description of the $Y_{n+1}$-tile that it simulates. We can now refer to this word as the state of the $Y_{n+1}$-tile.

The letters $0$ and $1$ are used for the binary encoding, $/$ is used to separate different \dfn{subfields} of the simulated tile, and $\sharp$ is used as an endmarker. The $Y_{n+1}$-tiles have the same structure as the $Y_n$-tiles. Hence, the simulated tile will be divided into subwords separated by the $/$ symbol, and each subword will contain the information of a field of the simulated tile\ie the simulated tile will have the form $\info.\lev/\info.\addr/\ldots/\info.\checkk\sharp\sharp\ldots\sharp$ (again keeping in mind that the fields \info.\addr, \info.\age\etc\ cannot be read from a single $Y_n$-tile, but rather they are written letter by letter on a segment of tiles).
    
At this point, note that we must have $\log{|Y_{n+1}|} \le B_n$ in order for the states of $Y_{n+1}$ to be represented with words of length $B_n$.

\item[\lmail, \rmail:]These fields will be used to exchange information between neighboring $Y_n$-macrotiles. They have the same alphabet $\Gamma$ as the \info\ field. The \lmail\ field will send the information of the simulated tile to the left-neighboring $Y_n$-macrotile and the \rmail\ field to the right neighbor. In this way, every macrotile will learn the simulated tile of its neighboring macrotiles.


\item[\prog:]This field contains the description, encoded over alphabet $\{0,1\}$, of an algorithm that, given as input $n$ and the states of three $Y_n$-tiles, outputs the unique tile that can be placed above them in $Y_n$ (or rejects if there is none).
At this point lies the heart of the construction:
the program written in the \prog\ field is a uniform program that governs the behavior of all $Y_n$.


\item[\work:]This field will be used to store intermediate data during the computation. 

\item[\checkk:]This field will be used as an input for the TM simulation by the macrotile. It is devoted to containing the character that is present in the layer of $\D_{A^\Ns}^*$ at the bottom-left corner of the macrotile.
For $n=1$ we require that the \checkk\ field of a tile is equal to the letter in the layer $\D_{A^\Ns}^*$.
\end{itemize}

As we have already pointed out, any configuration of $Y_1$ will have a very strong hierarchical structure: it can be divided into $Y_1$-macrotiles\ie rectangles of size $B_1\times T_1$ with coordinates $(0,0)$ on the lower left corner, 
that behave like $Y_2$-tiles, and the local constraints of $Y_2$ are satisfied between $Y_1$-macrotiles.
In particular, rectangles of size $B_1B_2 \times T_1T_2$ of $Y_1$-tiles will form $Y_2$-macrotiles which simulate $Y_3$ tiles. 
And so on: for every $n$, every valid configuration of $Y_1$ can be partitioned into \dfn{$Y_n$ macrotiles} of size $B_1 \cdots B_n \times T_1 \cdots T_n$ that simulate $Y_{n+1}$-tiles.

\subsection{Self-simulation}\label{s:details}

In this subsection, we will give further details about the construction of $Y_n$, and we will also describe how it simulates $Y_{n+1}$. The worktime of a $Y_n$-macrotile is divided in various subperiods during which an agent will perform different operations. These phases are the following:

\begin{enumerate}
\item \emph{Sending mail}:
During this period, the $Y_n$-macrotiles exchange their information, so that a macrotile learns the state of its left and right macrotile. To do this, we assume that at position $(0,0)$ there is an \dfn{agent} (a TM head that organizes the computation) that performs the following operation. It starts moving to the right until it meets the marker $\sharp$ in the \info\ field. Then, it turns to the left and starts reading the word backwards, at each step copying the letter that it reads onto the \lmail\ field, in which they move one step to the right at every time step. In this way, we will have a caravan of letters separated by one cell moving to the right in the \lmail\ field. It will take them $B_n$ steps to reach the right-neighboring macrotile. At the same time, when the agent reaches position $0$, it starts walking back to the right again until it reaches again the first $\sharp$ in the \info\ field. Since the length of the word in the \info\ field is of length $O(\log{B_n})$, the agent has enough time to reach this position before the \lmail\ from the neighboring macrotile starts coming in. When the first letter arrives, the agent stops it and moves one position to the left. Then, it stops the next letter that it meets and so on, until it eventually reaches position $0$. In this way, the \info\ field of the left-neighboring macrotile has been copied onto the \lmail\ field of the macrotile. This whole procedure has taken up time $B_n+O(\log{B_n}) \le 2B_n$. After that, the agent can organize a similar procedure to copy the mail of the right neighbor of a macrotile to the \rmail\ field. This takes another $O(B_n)$ steps.

An important remark is that the whole process does not depend on $n$. Namely, we can describe the action of the agent and of the letters moving in the \lmail\ and \rmail\ fields with a uniform program that works for every value of $B_n$.
    
\item \emph{Checking the level}: In this workperiod, we make sure that the \info.\lev\ is greater by $1$ than the \lev\ of the tiles that form the macrotile, in order to ensure that 
the $Y_n$-macrotile simulates a $Y_{n+1}$-tile. 

The way to achieve this is the following: our agent calculates the numbers in \addr\ (represented in binary) and in \lev\ (represented in unary). Then it checks that \info\ is $1$ if $\addr\le\lev+1$, and $/$ if $\addr=\lev+2$ (remember that \info.\lev\ is written first on the \info\ field). This whole comparison can be done by a TM independent from $B_n,T_n$ in time and space $O(\log{B_n}+n)=O(\log{B_n})$, assuming $B_n>>n$. Also, at the end of this workperiod, we can assume that the agent returns to the cell with address $0$.


\item \emph{Checking the coordinates}:
Now, every macrotile is conscious of the \info\ of its neighboring macrotiles. Recall that the word read in the \info\ fields of a macrotile has the form $\info.\lev/\info.\addr/\info.\age/\ldots/\info.\checkk\sharp\sharp\ldots\sharp$. The \info.\addr\ and \info.\age\ fields are binary representations of coordinates $i\in\co0{B_{n+1}}$ and $j\in\co0{T_{n+1}}$ respectively. These words are of length $\log{B_{n+1}},\log{T_{n+1}}=O(B_n)$. We will say that the address of a macrotile is $i$ when the macrotile has a binary representation of $i$ in its \info.\addr\ field.

 In the beginning of this workperiod, the agent reads the word $1^n$ written into the \lev\ field of the $Y_n$-macrotile. Then the agent writes down the binary representation of $B_{n+1}$ and $T_{n+1}$ in the \work\ field.
 A new restriction on how to chose $B_n$ and $T_n$ (see Subsection \ref{s:bntn}) is that, given $1^n$ as input, we can write down a binary representation of $B_{n+1}$ and $T_{n+1}$ in time polynomial in $n$.

    After that, the agent goes to the cells that encode \info.\addr\ (it can be found by reading \info\ until finding a $/$) and checks whether $\info.\addr=\lmail.\addr+1\bmod{B_{n+1}}$.
If they are not, then the tiling becomes invalid. This check can be done in time polynomial in $\log B_n$ and space $O(B_n)$. 
Similarly, the agent also checks that $\info.\age=\lmail.\age$. 

At the end of this worktime period, we have assured that the coordinates of the macrotiles are compatible. The operation in this level depends on $n$, but in a uniform way\ie there exists a fixed TM that, given $n$, outputs $B_n$ and $T_n$. This input can be read from \lev\ field.


\item \emph{Transmitting the \checkk\ field}:
During this workperiod, we want to make sure that the information of the \checkk\ field corresponds to the data in the layer $\D_{A^\Ns}$.
A local constraint already imposed this for the $Y_1$-macrotiles.
The other levels cannot perform the same (in order for the description to be uniform), but instead, the level can be transmitted between the levels. 
To do so, the agent goes to the cell with address $0$, memorizes the bit it sees in the \checkk\ field and then checks that the \info.\checkk\ field is equal to this bit. This can all again be done in time $O(\log{B_n})$. This process can be performed by a program independent of $n$.

The result of this process is that a $Y_n$-macrotile supported on the rectangle $\co x{x+B_n}\times\co y{y+T_n}$ carries in its \checkk\ field the character from the cell $(x,y)$ of the layer $\D_{A^\Ns}$.
We assume that $B_n$ is odd, so that (see the Point \ref{i:odd} of the remark in Subsection \ref{ss:dens}) the \checkk\ fields of the configuration of $Y_n$ and the configuration of $Y_{n+1}$ form two configurations $x,y\in\D_\alpha$ for some $\alpha\in\ans$ (they have the same densities).

\item \emph{Checking the input}: 
%
We can assume that at the beginning of this workperiod, the agent is in the cell with address $0$. During this workperiod, the agent performs $n$ steps of the TM $\m$, with input tape the \checkk\ field, and working tape the \work\ field. If $\m$ halts within $n$ steps, then the configuration is rejected (it is not in our final subshift).


\item \emph{Forcing self-similarity}: 
%
%
In this workperiod, we ensure that $Y_n$-macrotiles behave like $Y_{n+1}$-tiles. The behavior of $Y_n$-macrotiles is governed by the program written in \info.\prog, so we have to check that it is the same as the program written in \prog\ of every tile of $Y_n$. This is easily checked in the following way:
the agent 
goes to the cell where the first letter of the \info.\prog\ is held. Again, this can be done since the encoding of the information of the simulated tile in a macrotile is algorithmic and has a very explicit form. In this cell, it checks that the letter held in the \info\ field is equal to the first letter of the \prog\ field. Then, the agent goes one cell to the right and compares the letter in the \info\ field with the second letter of the \prog\ field, and so on, until a $/$ is read in the \info\ field. If at any cell the letters examined are not equal, the configuration is rejected.
This period takes $O(\log{B_{n+1}})$ steps.

\item \emph{Updating the state}:
The agent can apply the local rule encoded in \prog\ to the states described in the \lmail\, \info\ and \rmail\ fields. It does so by comparing this triple to the ones encoded in \prog. This will eventually give the result, which is written in the \info\ field bit by bit.
Then, the agent clears the \lmail, \rmail\ and \work\ fields. Fields \lev, \addr, \prog\ and \checkk\ are left unchanged; field \info.\age\ is incremented by $1$. This period also takes $O(\log{B_{n+1}})$ steps.
\end{enumerate}

\subsection{Choosing the right values for $B_n$ and $T_n$}\label{s:bntn}

In the previous construction, we have made some assumptions concerning the values of $B_n$ and $T_n$. First of all, let us notice that for every $n$, if $A_n$ is the alphabet of $Y_n$, we have $\card{A_n}\le O(nB_nT_n)$. Also, the whole computation contained in a $Y_n$-macrotile can be done in time $O(B_n)$, hence $T_n$ can be chosen to be of order $O(B_n)$. $B_n$ must satisfy the following restrictions:
\begin{itemize}
\item $\log{\card A_n}\le B_n$;
\item Binary representations of $B_n$ and $T_n$ are computable in time polynomial in $n$ with input given in unary;
\item $B_n$ is odd;
\item $B_n >> n$.
\end{itemize}
The first restriction is necessary in order to be able to represent $Y_{n+1}$-tiles with $Y_n$-macrotiles. The second one so that $T_n$ can be chosen relatively small compared to $B_n$. The third one is necessary so that we can read $\alpha$ from the sequence of macrotiles.

If we choose $B_n= c_13^n$ and $T_n = c_2B_n$, where $c_1,c_2$ are fixed numbers and $c_1$ is sufficiently large so that the inequalities are satisfied for small values of $n$, then all of the restrictions are satisfied.

%% file: entropies.bbl
\begin{thebibliography}{10}

\bibitem{entrind}
{\v C}ulik, {II}, K., Hurd, L.P., Kari, J.:
\newblock The topological entropy of cellular automata is uncomputable.
\newblock Ergodic Theory \& Dynamical Systems \textbf{12}(2) (1992)  255--265

\bibitem{entrss}
Simonsen, J.G.:
\newblock On the computability of the topological entropy of subshifts.
\newblock Discrete Mathematics \& Theoretical Computer Science \textbf{8}
  (2006)  83--96

\bibitem{entrsft}
Hochman, M., Meyerovitch, T.:
\newblock A characterization of the entropies of multidimensional shifts of
  finite type.
\newblock Annals of Mathematics \textbf{171}(3) (2010)  2011--2038

\bibitem{projsft}
Hochman, M.:
\newblock On the dynamics and recursive properties of multidimensional symbolic
  systems.
\newblock Inventiones Mathematic{\ae} \textbf{176}(1) (April 2009)  131--167

\bibitem{nilpind}
Kari, J.:
\newblock The nilpotency problem of one-dimensional cellular automata.
\newblock SIAM Journal on Computing \textbf{21}(3) (1992)  571--586

\bibitem{gacs}
G\'acs, P.:
\newblock Reliable cellular automata with self-organization.
\newblock Journal of Statistical Physics \textbf{102}(1--2) (2001)  45--267

\bibitem{appendix}
Guillon, P., Zinoviadis, C.:
\newblock Densities and entropies in cellular automata.
\newblock see the appendix below (2012)

\bibitem{milnor}
Milnor, J.:
\newblock On the entropy geometry of cellular automata.
\newblock Complex Systems \textbf{2}(3) (1988)  357--385

\bibitem{rhierarchy}
Zheng, X., Weihrauch, K.:
\newblock The arithmetical hierarchy of real numbers.
\newblock In Kutyłowski, M., Pacholski, L., Wierzbicki, T., eds.:
  Computer Science ( MFCS'99. Volume 1672 of LNCS., Springer Berlin/Heidelberg
  (1999)  23--33

\bibitem{coneentr}
Park, K.K.:
\newblock Entropy of a skew product with a $\mathbbm z^2$-action.
\newblock Pacific Journal of Mathematics \textbf{172}(1) (1996)  227--241

\bibitem{robinson}
Robinson, R.M.:
\newblock Undecidability and nonperiodicity for tilings of the plane.
\newblock Inventiones Mathematic\ae \textbf{12}(3) (1971)

\bibitem{fixpttile}
Durand, B., Romashchenko, A., Shen, A.:
\newblock Fixed-point tile sets and their applications.
\newblock draft (September 2010)

\bibitem{nexpdir}
Hochman, M.:
\newblock Expansive directions for $\mathbbm z^2$ actions.
\newblock Ergodic Theory \& Dynamical Systems \textbf{31}(1) (2011)  91--112

\bibitem{mozes}
Mozes, S.:
\newblock Tilings, substitution systems and dynamical systems generated by
  them.
\newblock Journal d'analyse math\'ematique \textbf{53} (1988)  139--186

\bibitem{twobytwo}
Ollinger, N.:
\newblock Two-by-two substitution systems and the undecidability of the domino
  problem.
\newblock In Beckmann, A., Dimitracopoulos, C., L\"owe, B., eds.: 
  theory of algorithms, Computability in Europe ( CiE'2008. Volume 5028 of
  LNCS., Athens, Greece, Springer Berlin / Heidelberg (June 2008)  476--485

\bibitem{expsubd}
Boyle, M., Lind, D.:
\newblock Expansive subdynamics.
\newblock Transactions of the American Mathematical Society \textbf{349}(1)
  (1997)  55--102

\end{thebibliography}
